# Design of a Modified Coupled Resonators Optical Waveguide Supporting a Frozen Mode

Mohamed Y. Nada[1], Tarek Mealy[1], Md Shafiqul Islam[1], Ilya Vitebskiy[2], Ricky Gibson[2], Robert Bedford[3], Ozdal Boyraz[1], and Filippo Capolino[1]

*(1) Department of Electrical Engineering and Computer Science, University of California, Irvine, CA 92697, USA*

*(2) Air Force Research Laboratory, Sensors Directorate, Wright-Patterson Air Force Base, OH, 45433, USA*

*(3) Air Force Research Laboratory, Materials and Manufacturing Directorate, Wright-Patterson Air Force Base, OH, 45433, USA*

We design a three-way silicon optical waveguide with the Bloch dispersion relation supporting a stationary inflection point (SIP). The SIP is a third order exceptional point of degeneracy (EPD) where three Bloch modes coalesce forming the frozen mode with greatly enhanced amplitude. The proposed design consists of a coupled resonators optical waveguide (CROW) coupled to a parallel straight waveguide. At any given frequency, this structure supports three pairs of reciprocal Bloch eigenmodes, propagating and/or evanescent. In addition to full-wave simulations, we also employ a so-called "hybrid model" that uses transfer matrices obtained from full-wave simulations of sub-blocks of the unit cell. This allows us to account for radiation losses and enables a design procedure based on minimizing the eigenmodes' coalescence parameter. The proposed finite-length CROW displays almost unitary transfer function at the SIP frequency, implying a nearly perfect conversion of the input light into the frozen mode. The group delay and the effective quality factor at the SIP frequency show an $N^3$ scaling, where $N$ is the number of unit cells in the cavity. The frozen mode in the CROW can be utilized in various applications like sensors, lasers and optical delay lines.

## I. Introduction

An exceptional point of degeneracy (EPD) in a system parameters space is the point at which two or more system eigenmodes coalesce in both eigenvalues and eigenvectors [1]–[5]. The EPD has a degeneracy order that is determined by the number of coalescing eigenmodes. Although most of the published work on exceptional points is based on PT symmetry [3], [4], the occurrence of an EPD actually does not require a system to satisfy PT symmetry. Indeed, EPDs have been recently found also in single resonators by just adopting time variation of one of its components [6].

In this paper, we focus on an EPD of third order in a periodic photonic structure. At such a point, one propagating and two evanescent Bloch eigenmodes collapse on each other forming the frozen mode (see, for example, [2], [7]–[9] and references therein). The dispersion relation of the propagating component of the frozen mode develops a stationary inflection point (SIP) at the EPD frequency. The most prominent feature of a SIP supporting periodic structure is the frozen mode regime, which is a conversion of the input signal at the respective frequency into a slow mode with greatly enhanced amplitude. Applications of the frozen mode regime include but are not limited to pulse compressors [7], optical memory devices, antennas, filters [10], optical switching [11], lasers [8], [12], and tunable optical delay lines [13]. Moreover, the SIP slows down the electromagnetic waves to allow strong light matter interaction, which can be used to increase the wall plug efficiency of lasers or to obtain high-gain high-power amplifiers [9].

The SIP was found in non-reciprocal structures [2], [8], [14] using magnetic materials to break the system reciprocity. It has been shown that the SIP can occur also in lossless, reciprocal, structures, made of a three-way waveguide, i.e., a waveguide that supports three modes in each direction. Such three-way waveguides can be made of optical coupled silicon ridge waveguides [15], optical coil resonators [16], [17], the modified coupled resonator optical waveguide (CROW) [5], and also using a serpentine optical waveguide [18].

In this paper, we present a CROW-based design of an optical three-way waveguide that exhibits SIP. The fundamental idea is similar to the one in [5] but it is here realized in Si on insulator (SOI) technology. The proposed structure exhibits an SIP; earlier SIP demonstrations in optical waveguides were either based on coupled mode theory [5], [18], [19] or on holed ridge waveguides [20], [13], [21], [15] with circular or rectangular holes. The SOI platform has emerged as a promising technology for realizing photonic integrated systems. This platform offers low loss passive photonic components as well as a wide range of active components. Mature fabrication process and CMOS compatibility are two key factors that has attracted widespread attention to SOI platform.

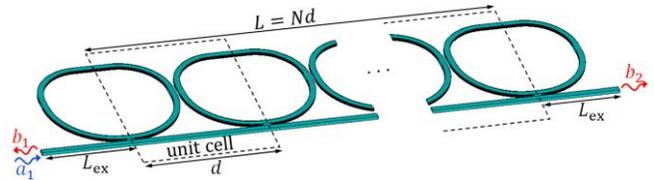

FIG. 1. Finite-length modified CROW coupled to a parallel straight waveguide. The CROW structure forms a cavity made of $N$ unit cells terminated by completing the racetrack resonators and extending the straight waveguide by length $L_{ex}$. The conventional CROW is modified by deforming the ring shape by introducing two radii and by adding straight waveguide sections. The radii in the upper two ring quarters different than the radii of the lower ones. Note that the whole structure is made of $N+1$ modified rings and the total length of the straight waveguide is $L+2L_{ex}$.

The designed three-way waveguide is made of a CROW that is longitudinally coupled to a straight waveguide, and we show that the proposed design exhibits an SIP as it was originally proposed in [5] using coupled mode theory, with the difference

This material is based upon work supported by the Air Force Office of Scientific Research award numbers LRIR 21RYCOR019 and FA9550-18-1-0355.. Mohamed Y. Nada and Tarek Mealy contributed equally to this work.

that in this paper the unit cell has a *single* non-circular ring whereas in [5] the unit cell had *two* rings that are circular.

This paper also presents an original and accurate "hybrid" method to design complex optical waveguides, that is much more prone to be used for optimization than full-wave simulations of the whole unit cell.

In Sec. II, we discuss the CROW unit cell made of only one ring where we have introduced a new design degree of freedom compared to the CROW proposed in [5] by making the radii in the upper two ring quarters of the modified ring resonator different from the lower ones. In Sec. III, we present the "hybrid" model used to design the CROW and discuss its accuracy and the optimization time. Indeed, these CROW waveguides are very large in terms of optical wavelengths, and smart schemes for modeling them are needed to preserve processing time of the unit cell dimensions, based on the coalescence of three eigenvectors that are computed numerically. Sec. IV, shows the optimized CROW dimensions that lead to an SIP. In Sec. V, we show the dispersion diagram of the eigenmodes of an infinitely-long periodic CROW, and we show that it exhibits an SIP. In Sec. VI, we explore the properties of the finite-length CROW cavity operating near the SIP, show the transfer function, the reflection coefficient, the quality factor, and the group delay based on full-wave simulations. In Sec. VII, we show the effect of structural perturbations on the occurrence of the SIP.

## II. Geometry of the SIP-CROW

In this paper we show a practical design of the three-way CROW proposed in [5] so that we realize the third order EPD, i.e., the SIP. A brief theoretical study of the SIP in CROWs was presented in [5] using couple mode theory but in this paper we deepen our study about SIP in optical resonators and we verify the existence of the SIP in a novel geometry with practical dimensions through full-wave simulations.

The unit cell of the CROW proposed in [5] was designed to realize an SIP with two rings in the unit cell where the coupling coefficients are different in the two rings. A thorough discussion is presented in [5] about why such a unit cell was chosen. In this paper we propose a new cavity design shown in Fig. 1 based on a more compact unit cell consisting of only one ring as shown in Fig. 2(a). The cavity is made of a chain of coupled racetrack resonators, each involving two different radii $R_b$ and $R_t$ as shown in Fig. 2(a). The coupling in the proposed racetrack CROW is realized by directional couplers, i.e., the coupling in Fig. 1 is distributed rather than point coupling as in [5]. The coupling between two adjacent rings is determined by a directional coupler of length $L_{rr}$ while the gap in the couplers is $g$, hence we call the ring resonator as a racetrack resonator. The chain of rings is side coupled to a uniform optical waveguide through a directional coupler of length $L_{wr}$ and gap $g$, similar to the gap between the rings. The upper horizontal flat part of the racetrack resonator is related to the radii and $L_{wr}$ such that it satisfies

$$L_t = L_{wr} + 2R_b - 2R_t. \qquad (1)$$

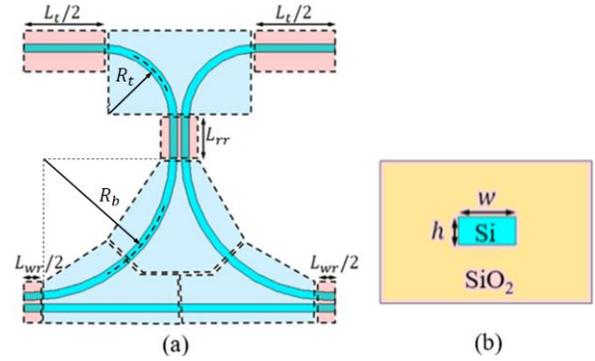

FIG. 2. (a) Unit cell of the modified CROW waveguide shown in Fig. 1, where the radii of the top and bottom quarter rings are different from each other. We divide the unit cell into sub-blocks where the transfer matrices of blue-shaded sub-blocks are obtained directly from 3D full-wave simulations whereas the ones of the red shaded sub-blocks are obtained based on transmission line model using phase velocities and characteristic impedances calculated from 2D full-wave port analysis simulations. The coupling between the straight waveguide and the ring is dictated by a directional coupler of length $L_{wr}$, while the coupling between the rings is dictated by a coupler of length $L_{rr}$. All couplers are assumed to have the same gap $g$. The horizontal segment on the top with length $L_t = L_{wr} + 2R_b - 2R_t$ is added to complete the racetrack. (b) Cross section of the Si waveguides width $w$ and height $h$. The waveguide is surrounded by a cladding of silicon dioxide.

Such a modification of introducing two different radii in the racetrack resonator adds an extra degree of freedom to the design, than what was presented in [5]. The single-ring unit cell with two radii simplifies the design process of the CROW so that it may exhibit an SIP in the dispersion diagram. The unit cell of the CROW shown in Fig. 2(a) has a period $d = 2R_b + L_{wr} + w + g$. The waveguides are chosen to be SOI strip waveguides as shown in Fig. 3(a). The fully etched strip waveguide provides tight confinement due to the high index contrast between the silicon core and the SiO$_2$ cladding. The rest of the analysis in this paper uses the refractive indices of $n_{si}$=3.48 and $n_c$=1.45 for the silicon core and the glass cladding, respectively, as shown in Fig. 2(b). Moreover, the waveguide is designed to have a height of $h = 220$ nm and width of $w = 450$ nm. The dimensions are chosen to ensure single transverse electric (TE) mode operation (i.e., with electric field along the horizontal direction) as shown in Fig. 3(b). In Fig. 3(c), we show the minimum width for each waveguide height that enables multimode operation (without considering the TM mode), which confirms that choosing $w = 450$nm at a height 220nm guarantees the single mode operation. Waveguide with similar height is common for MPW (Multi-Project Wafer) services offered by different foundries such as the Interuniversity Microelectronics Centre (IMEC) [22], the American Institute for Manufacturing (AIM) Photonics [23].

We performed numerical mode calculations for the waveguide structure shown in Fig. 2(b). We increased the waveguide width with 10 nm steps for any fixed waveguide height. Our goal is to determine the waveguide dimensions that would allow a higher-order TE mode to propagate. In general, numerical simulations provide ideal propagating modal solutions that however, in practice, would not "survive", i.e., they would be indeed attenuated, due to roughness, imperfections, and curvature. There are no clear-cut established criteria for deciding which mode survives in a practical device. Here, we define a mode as propagating and surviving when the effective mode index is higher than the cladding refractive index, and the mode confinement factor (cf) is above some threshold value. The latter condition helps rule out poorly confined modes whose effective index is very close to the cladding refractive index and will not propagate in an actual device. Those higher-order modes will radiate away at slight micro bends, such as in the resonators used in this paper. Figure 3(c) shows borderlines (i.e., thresholds) of waveguide dimensions at which the multimode propagation starts, for confinement factors thresholds varying from 20% to 35%. A waveguide with width and height $(w, h)$ lying on the right of a depicted curve will satisfy cf above the selected threshold. Since the width is increased by 10 nm discrete steps, an error bar corresponding to 10 nm step size is added to show the uncertainty of results in Fig. 3(c). Vanishing error bars in Fig 3(c) corresponds to points that generate the exact value of the selected confinement factor (cf). The detailed simulation settings can be found in Appendix A. For practical purposes we could consider a mode non surviving if the cf is less than 35%, though this threshold is somewhat arbitrary.

### III. Hybrid Model of the SIP-CROW

Throughout the paper we assume that the time convention is $e^{j\omega t}$. We define a state vector at the middle of the waveguide-ring coupler, representing the fields at the boundaries of the unit cell in Fig. 2. The state vector represents the electric field and magnetic field components at each unit cell as

$$\boldsymbol{\Psi}_n = [V_{e,n}, \ I_{e,n}, \ V_{o,n}, \ I_{o,n}, \ V_{3,n}, \ I_{3,n}]^T \quad (2)$$

where $n$ is the unit-cell index, $V_{e,n}$, $V_{o,n}$, $I_{e,n}$ and $I_{o,n}$ are the equivalent voltages and currents representing the coupled transmission lines (CTL) model of the coupled waveguide, and $V_{3,n}$ and $I_{3,n}$ are the equivalent voltage and current representing the transmission line (TL) model of the uncoupled waveguide. We present in Appendix B and Appendix C the way we define the voltages and currents for coupled and uncoupled waveguides. While in Appendix D, we present the way we obtain the transfer matrices for the pink-shaded blocks based on S-matrices that are found using full-wave simulations based on the finite element method.

The eigenvalue problem and the dispersion diagram are obtained following the analysis presented in [5], [24], [25] that was based on coupled mode theory [26], [27]. Hence, we first evaluate the 6×6 transfer matrix (T-matrix) $\underline{\mathbf{T}}$ representing the evolution of the state vector in (2) across a unit cell and the dispersion diagram is then obtained though the dispersion relation

$$D(k, \omega) \equiv \det(\underline{\mathbf{T}} - \zeta \underline{\mathbf{1}}) = 0 \quad (3)$$

where the eigenvalues are $\zeta = \zeta_n \equiv e^{-jk_n d}$, with $n = 1, 2, ..., 6$ and $\underline{\mathbf{1}}$ is the 6×6 identity matrix. In this paper, the transfer matrix is obtained numerically, not calculated theoretically as in [5], from the scattering parameters resulting from the full-wave simulations of the unit cell in Fig. 2(a).

To obtain an SIP, one needs to optimize the unit cell dimensions that involves obtaining the eigenvectors of $\underline{\mathbf{T}}$ at each step, and changing the dimensions of the geometry until three eigenvectors coalesce or become very close to each other. This is a challenging task because the radii and the size of the unit cell are usually in few tens of µm which is huge compared to the gap size that is typically few tens of nm. To have accurate results from full-wave simulations, the mesh has to be fine which take quite some time to simulate even one unit cell. To get sense of the numbers, in order to simulate one unit cell of the SIP CROW (will be shown later), the number of mesh cells is around $2 \times 10^6$ tetrahedrons and the full-wave simulation time is around 6 hours on a machine with 128 GB RAM and 2 processors of 2.6 GHz speed. This makes the optimization process through full-wave simulations almost impossible as it requires very long time to make iterations of simulations in order to optimize the unit cell parameters to get an SIP. To solve this issue and make the optimization a realistic process, we use a model for the unit cell that takes short time to simulate and closely matches the full wave simulation results, which we refer to as the *hybrid model*.

Transmission line models are used to model the coupled and uncoupled sections of waveguides using the transfer matrices in Appendix C. For accurate modeling, the impedance and phase velocity of the propagating mode in a single straight waveguide, and those for even and odd modes of coupled straight waveguides are calculated using 2D full-wave port analysis solver implemented in CST Studio Suite as shown in Appendix C. Such values are needed to build the T-matrices as shown in Appendix C for a single TL and coupled TLs.

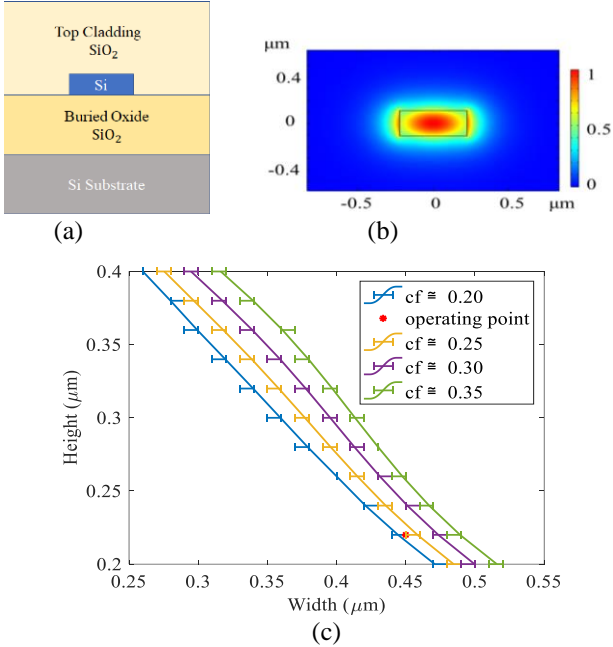

FIG. 3. (a) Silicon strip waveguide geometry with SiO$_2$ cladding. (b) Normalized |**E**| field pattern for the fundamental TE mode of the Si strip waveguide. (c) Curves with constant confinement factor (cf) versus width and height of the strip Si waveguide. Choosing a level of cf (i.e., 35%) defines the maximum dimensions of the strip that guarantees a single TE mode. The error bars on data points indicate uncertainty due to width step used in the numerical search of single mode operation. The red asterisk marks the dimensions of the waveguide cross section used in this paper.

A unit-cell model that is fully based on TLs could be used to optimize the dimensions of the unit cell in Fig. 2(a) by modeling the curved waveguides as single TLs and by accounting for their proper path lengths. However, such a model ignores important factors like the coupling between the waveguides in the curved areas, the slight change in the effective refractive index (or equivalently the change in impedance and phase velocity) due to the curvature, and radiation losses. Therefore, the use of such a simple TL model is not a good candidate for optimization to get an SIP because it will give results that is not accurate, i.e., far from the actual ones related to the real geometry.

Hence, we use a hybrid model that has the advantages of very good accuracy and moderate optimization time. We divide the unit cell into different sections as shown in Fig. 2(a) and use the TL model to calculate the T-matrices of the single waveguides and the directional couplers (highlighted by pink color in Fig. 2(a)). On the other hand, the T-matrices of the other sections that includes curvatures (highlighted by light blue color in Fig. 2(a)) are calculated by converting the S-parameters obtained from full-wave simulations implemented based on the finite element method using CST Studio Suite. The T-matrix of the whole unit cell is then calculated by cascading the T-matrices of the different sections in the proper order. In full-wave simulations, the polarization of the port modes, upon which the S-matrix is built, may have $180^o$ phase shift as their phases are selected arbitrarily by the mode solver. Therefore, such inconsistency of modes' polarizations should be carefully taken care of to properly concatenate the S-matrices. In this hybrid model we fix the gap $g$ and the radii $R_b$ and $R_t$ leaving the couplers' lengths to be optimized using MATLAB, i.e., we have two variables to optimize $L_{wr}$ and $L_{rr}$ in the unit cell shown in Fig. 2(a).

We validate the accuracy of the hybrid model for the proposed unit cell by comparing the dispersion diagram of the eigenmodes in an infinitely long CROW obtained from CST full-wave simulations of the whole unit cell with the dispersion diagram based on the hybrid model. The dispersion diagram comparison is shown in Appendix E for arbitrary coupler lengths, where we can clearly see the very good agreement between the results. Note that the dispersion diagram does not show an SIP since this validation step comes before the optimization process.

To summarize, three methods could be used to obtain the T-matrix and optimize the unit cell dimensions: (i) the full-wave simulations of the whole unit cell implemented in CST, though this is slow (in terms of a single unit cell simulation time) and cannot be easily used to optimize the CROW parameters, (ii) the TL-model which is fast and very easy to optimize but it lacks accuracy as it is based on several approximations, and (iii) the " hybrid model" that provides good speed, and it offers advantages in both accuracy and flexibility for the optimization process to obtain the SIP. One could even argue that this hybrid method is at least as accurate as the full-wave method applied to the whole unit cell since the computation domains are smaller.

## IV. Design of Unit Cell Dimensions and the Optimization Process

After having a valid model of the CROW unit cell, we come to the choice of the design dimensions and the optimization process. We discuss here the optimization of the unit-cell parameters based on the hybrid model.

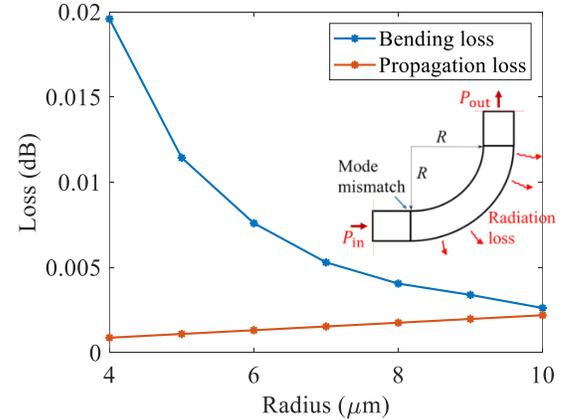

FIG. 4. Bending loss of 90° bend silicon waveguide vs. radius. It describes radiation from the curved part and the two sections where the waveguide changes radius of curvature (inset shows the simulation setup). The propagation loss of a straight or curved waveguide (due to material loss and roughness, without considering bend losses) for the same 90° bend silicon waveguide is also shown for comparison.

As explained, in the hybrid model, we fix the gap between the couplers $g$, and the radii $R_t$, $R_b$, whereas we are left with two parameters to be optimized $L_{wr}$ and $L_{rr}$. The choice of the ring radii is a tradeoff between the propagation losses and radiation losses. The EPD is a precise condition that is very sensitive to perturbation, and losses affect it (though they can be in part compensated by the presence of gain). Hence, our goal is to optimize the design of the passive waveguide trying to minimize the effect of losses. The loss in a straight waveguide using SOI as shown in Fig. 3(a) can be less than 1.4 dB/cm [22], [28] as reported by the Interuniversity Microelectronics Centre (IMEC) . The loss recorded by the American Institute for Manufacturing (AIM) for a straight waveguide is less than 2.5 dB/cm [23], [29]. Also due to tight mode confinement, an SOI strip waveguide bend has very low radiation losses. For example, in [30] the authors reported an experimental bend loss of 0.005 dB/90° for 5 µm bend at 1500 nm, where the waveguide was 445nm wide and 220 nm high. In [31], a bend loss of 0.009dB/90° is reported for a 500 nm× 220nm waveguide at 1550 nm. Smaller radii imply larger radiation losses; however, using larger ones means more propagation losses, and also more ring resonances packed together in frequency, i.e., smaller free spectral range (FSR).

We have simulated bending loss for a 90° bend using the setup shown in Fig. 4. We define loss as $L_{\mathrm{dB}} = P_{\mathrm{in,dB}} - P_{\mathrm{out,dB}}$. However, full-wave simulations provide the scattering parameter $|S_{21}|^2 = P_{\mathrm{out}}/P_{\mathrm{inc}}$, where $P_{\mathrm{inc}}$ is the incident power. The input power is written in terms of the incident power as $P_{\mathrm{in,dB}} = 10 \log(P_{\mathrm{inc}}(1 - |S_{11}|^2)) = P_{\mathrm{inc,dB}} + 10 \log(1 - |S_{11}|^2)$. From simulation, we found that $S_{11}$ is in the order of $-50$ dB (i.e., $|S_{11}|^2 \ll 1$). Therfore, the input power is approximated (using the Taylor series for log function) to $P_{\text{in,dB}} \approx P_{\text{inc,dB}} - 10|S_{11}|^2 \log e$. Thus, under the assumption that $|S_{11}|^2 \ll 1$, the loss is approximated to $L_{\text{dB}} \approx -(P_{\text{out,dB}} - P_{\text{inc,dB}}) - 10|S_{11}|^2 \log e \approx -10 \log(|S_{21}|^2)$. The loss for the case under study is resulting from mode mismatch loss (i.e., radiation from the radius of curvature discontinuity) and radiation loss due to the bent waveguide itself.

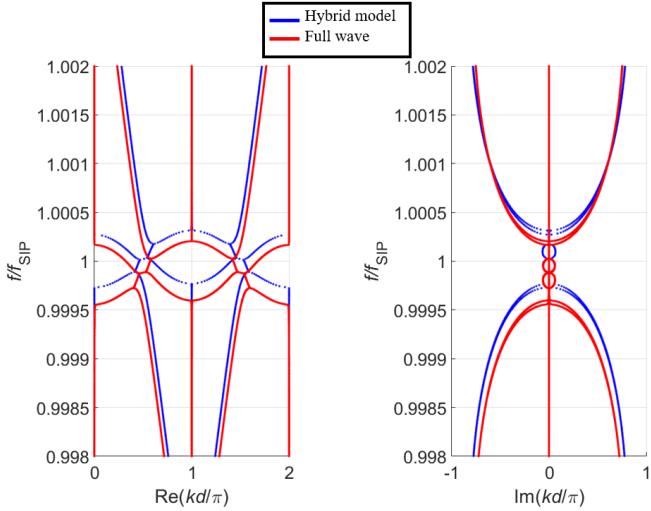

FIG. 5. Complex dispersion diagram of the modes in a CROW with unit cell made of one ring shown in Fig. 2(b). The SIP is at a frequency $f_{\text{SIP}} = 193$ THz corresponding to the optical wavelength of $\lambda = 1550$ nm. Dimensions of the unit cell: gap $g$=200nm, $R_t$=10μm, $R_b$=11μm, $L_{rr}$=0.365μm, and $L_{wr}$=0.273μm. The dispersion diagram obtained from the hybrid model is in a perfect agreement with the full-wave simulations except for a small frequency shift.

In Fig. 4, we show the bending loss in dB for a 90° bend silicon waveguide calculated for different radii. The loss is calculated here as $L_{\text{dB}} = -10 \log(|S_{21}|^2)$. The scattering parameters are obtained from full-wave simulation based on the the finite element method implemented in CST Studio Suite. A bend radius $R = 10$ μm gives a bend loss of 0.0026 dB/90°. We also compare the propagation loss due to scattering caused by fabrication imperfections with the simulated bending loss. For a 90° bend with bend radius $R$, we calculate the propagation loss (without bend-induced radiation loss) as the product of the propagation loss in dB/cm and the length of the quarter ring $\pi R/2$ in cm where we use the value of propagation loss of 1.4 dB/cm as reported in [22] for a waveguide with the same geometry and dimensions as the one considered here. Figure 4 shows the propagation loss in the curved 90° bend for different radii. Propagation losses due to imperfections are always lower than radiation losses for almost all radii considered, and they are comparable for $R = 10$ μm. We choose the radii $R_t$ and $R_b$ of the design shown in Fig. 2(a) to be in the order of 10μm as an optimized value to get the minimal radiation and propagation losses.

After choosing the radii, we optimize the couplers' lengths by minimizing the coalescence parameter $C$ which is a figure of merit to evaluate how close the system is to the exact EPD condition.

The coalescence parameter was presented in [32], [33] for the EPDs of order 4 and 3, respectively. The coalescence parameter in the case of EPD of order 3, the SIP, is a measure of the three-dimensional angle between the eigenvectors of the eigenmodes existing in the structure. Following [33], the coalescence parameter $C$ for the three coalescing eigenvectors is calculated as

$$C = \frac{1}{3}\sum_{\substack{m=1,n=2 \\ n>m}}^{3}|\sin\theta_{mn}|, \cos(\theta_{mn}) = \frac{|\langle\Psi_m,\Psi_n\rangle|}{\|\Psi_m\|\|\Psi_n\|}, \quad (4)$$

where $\Psi_n$, with $n = 1,2,3$, are the three six-dimensional normalized complex eigenvectors of the eigenmodes with wavenumbers $0 < \text{Re}(kd/\pi) < 1$. Furthermore, $\theta_{mn}$ is the angle between the two eigenvectors $\Psi_m$ and $\Psi_n$, and it is defined via the inner product

$$\langle\Psi_m,\Psi_n\rangle = \Psi_m^\dagger \Psi_n. \quad (5)$$

The dagger symbol † represents the complex conjugate transpose operation, and $\|\Psi_m\|$ and $\|\Psi_n\|$ denote the norms. The coalescence parameter $C$ is always positive and less than one, and $C = 0$ indicates the perfect coalescence of the three eigenvectors, i.e., the system experiences an SIP. Hence, the optimum lengths of the couplers where the system is as much as close to the SIP are those at which $C$ is minimum.

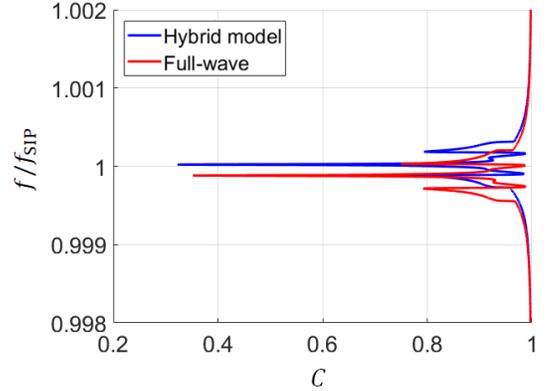

FIG. 6. Coalescence parameter calculated based on the hybrid model and full-wave simulations, where the minimum value represents the coalescence of three eigenvectors, i.e., three eigenvectors are almost parallel.

## V. Dispersion Diagram of Eigenmodes in SIP CROW

We show in Fig. 5 the complex dispersion diagram of the eigenmodes in an infinitely-long periodic CROW whose unit cell is made of one racetrack resonator as shown in Fig. 2(a), i.e., with waveguide dimensions of $h = 220$ nm and $w = 450$ nm. The refractive indices of silicon and silicon dioxide are considered as 3.48 and 1.45, respectively. The other dimensions are $g = 200$ nm, $R_t = 10$ μm, $R_b = 11$ μm. The coupling lengths are $L_{rr} = 0.365$ μm leading to the magnitude of the field coupling coefficient between the rings of approximately 2.4 %, and $L_{wr} = 0.273$ μm leading to a field coupling coefficient of 1.8 % between the rings and the straight waveguide. In Fig. 5, we show both the dispersion diagram obtained based on the hybrid model represented by blue lines and the dispersion obtained from the full-wave simulation of the unit cell implemented fully in CST Studio Suite represented by red lines. The two models are in perfect agreement except for a slight and acceptable shift in frequency.

For the hybrid model, the pink shaded blocks in Fig. 2(a) are modeled via single and coupled TLs with phase velocities and impedances reported in Appendix C. The blue shaded blocks are directly modeled based on S matrices (See details in Appendix D) obtained from the finite element method implemented in CST Studio Suite using adaptive mesh. The simulations stopping criterion has thresholds of $5 \times 10^{-5}$ and $2 \times 10^{-4}$ for the error of the port mode normalized wavenumber $k_z/k_0$ and the maximum deviation of the absolute value of the complex difference of the S-parameters, respectively, between two subsequent passes.

In the dispersion diagram, the SIP is characterized by $\omega_e$ and $k_e$, and in its vicinity the dispersion is well approximated by

$$(1 - \omega/\omega_e) \approx \eta_e (1 - k/k_e)^3, \qquad (6)$$

where $k_e$ is the SIP wavenumber at the exceptional point. In analogy to the theory presented in [24], the dimensionless "flatness parameter" $\eta_e$ is related to the third derivative of $\omega$ with respect to $k$ at the SIP angular frequency $\omega_e$, i.e.,

$$d^3\omega/dk^3 = -\eta_e \omega_e / (6 k_e^3), \qquad (7)$$

and it dictates the flatness of the dispersion relation at $\omega_e$. The three complex branches coalescing at the SIP frequency associated to Eq. (6) can be seen in Fig. 5, where one branch has almost a purely real wavenumber.

The dimensions and parameters of the unit cell have been optimized based on the coalescence parameter $C$. However, in the dispersion diagram in Fig. 5, the SIP is not ideal, i.e., there is no perfect coalescence as discussed next based on the concept of coalescence parameter. In Fig. 6, we show the eigenvectors coalescence parameter using both the hybrid model and the full-wave simulations, where the minimum value refers to the SIP frequency. This value ideally should be zero, but it is slightly greater than zero because we consider here losses that prevent the complete coalescence of the eigenmodes. Again, the two methods in Fig. 6 are in a perfect agreement except for the small frequency shift observed also in Fig. 5. Besides being used in the optimization process to find the dimensions that lead to an SIP, the coalescence parameter is also very helpful to assess how far are we from the ideal EPD condition [33].

*Effect of propagation losses*:

We consider a propagation loss of 2 dB/cm, that is modeled by assuming a lossy Si with a dielectric constant having $\tan(\delta) = 3 \times 10^{-6}$. To capture the change in the dispersion diagram of the structure with such very small loss, a very small threshold in the numerical calculations of the s-parameters admitted error should be used as stopping criterion for the adaptive mesh used in finite element method implemented in CST Studio Suite. This may require a super fine mesh with hundreds of millions of mesh cells which is not practical because of limited computer memory resources and because it would take a very long simulation run time. The change in the dispersion relation due to adding such small loss is expected to be almost negligible comparing to the simulation numerical error when a reasonable value for the stopping criterion threshold of the error in s-parameters of $2 \times 10^{-4}$ is used. Therefore, we expect that adding the propagation loss will make almost negligible changes to the dispersion relation without affecting the occurrence of the SIP.

The dispersion diagram in Fig. 5 also shows the regular band edges (RBEs) just above and below the SIP frequency. The existence of these RBEs cannot be avoided because of the branch representing two complex conjugate wavenumbers that have to bifurcate at some frequency. Those RBEs are not favorable here because they may interfere with the SIP in applications like laser when an active medium with broad bandwidth gain profile is used. Multiple simulations were performed to find possible ways to push those RBEs away from the SIP. We found that this can be achieved by either narrowing the coupling gap between the two waveguides or by reducing the ring perimeter by multiple of wavelengths calculated at the SIP frequency.

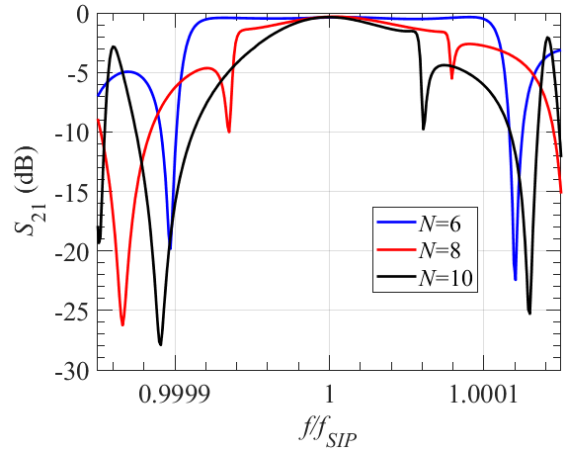

FIG. 7. Magnitude of the transfer function $S_{21}$ in dB of an SIP-CROW cavity operating in close proximity of the SIP frequency calculated using full-wave simulations implemented in CST Studio Suite. The TF is calculated for three different number of unit cells (*N*) of the SIP-CROW. The transfer function is almost unity at the SIP frequency.

## VI. CROW Cavity with SIP

We investigate in this section some properties associated with a CROW cavity shown in Fig. 1 operating near an SIP. We start by showing the transfer function of a cavity made of a lossless finite-length CROW as a function of frequency (*f*) for different numbers of unit cells *N*. A unit cell starts at the center of the ring, just in the middle of the directional coupler coupling the straight waveguide to the ring, as shown with a dashed line in Fig. 1. At each end, i.e., at $z = 0$ and $z = L$, there are three ports to be terminated: the first and last rings are terminated with half racetrack resonators, whereas the straight waveguide is extended for an extra length $L_{ex}$ without changing the waveguide cross-sectional dimensions. We define the input and output ports at $z = -L_{ex}$ and $z = L + L_{ex}$, with the extra length $L_{ex} = d/2 - g/2$. The input normalized wave amplitude $a_1$ is at the start of the straight waveguide (Fig. 1), where $b_1$ is the reflected normalized wave amplitude, on the other side the output normalized wave amplitude of the CROW is $b_2$. The transfer function is then defined by the scattering matrix coefficient $S_{21}$ as

$$S_{21}(f) = \frac{b_2}{a_1}. \quad (8)$$

The scattering parameters (power wave amplitudes) are defined based on a reference impedance that is the same as the wave impedance of a single waveguide $Z_{0w}$ obtained from the full-wave Port analysis (see details in Appendix B).

To calculate $S_{21}(f)$, we first obtain the state vector at the right boundary of the last unit cell, as $\underline{\Psi}(z = L) = \underline{\mathbf{T}}^N \underline{\Psi}_0$. Here $\underline{\mathbf{T}}$ is the T-matrix of one unit-cell and $\underline{\Psi}_0$ is the state vector defined at $z = 0$. Then, we apply the boundary conditions at both ends of the finite length CROW. The magnitude of the transfer function $S_{21}(f)$ is shown in Fig. 7 for three distinct numbers of unit cells $N$.

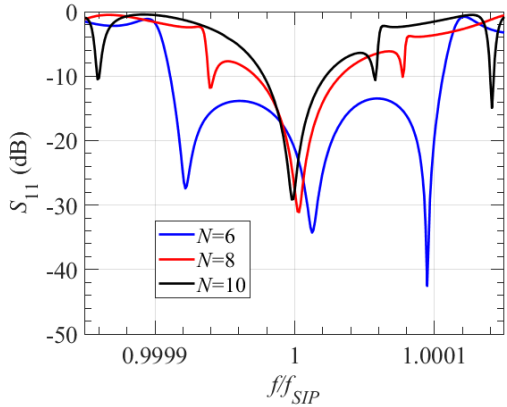

FIG. 8. Magnitude of the reflection coefficient $S_{11}$ in dB of the SIP-CROW cavity for different number of unit cells $N$. There is a good matching in proximity of the SIP frequency.

Note that the number of coupled racetrack resonators is $N+1$, as the unit cell starts in the middle of a resonator. It is clear from Fig. 7 that the transfer function is almost unity at the SIP frequency and that it does not show sharp resonances near the SIP.

In Fig. 8 and for the same three numbers of unit cells $N$, we show the reflection coefficient $S_{11}(f)$ defined as

$$S_{11}(f) = \frac{b_1}{a_1} \quad (9)$$

We see from Fig. 8 that the CROW cavity provides very good matching at the SIP frequency. Also, by increasing the number of unit cells we can get better matching exactly at the designed SIP frequency as we are approaching the ideal SIP condition defined for infinite structure.

An important application of CROW cavities operating near an SIP is the design of optical delay lines [13], [34], [35]. In Fig. 9, we investigate the group delay of the finite CROW cavity defined based on the $e^{j\omega t}$ time convention as the negative derivative of the transfer function phase with respect to the angular frequency,

$$\tau_g = -\frac{\partial \angle S_{21}(\omega)}{\partial \omega}. \quad (10)$$

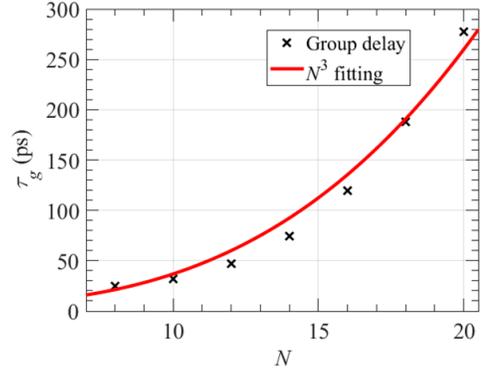

FIG. 9. Group delay of the CROW cavity shown in Fig. 1 calculated for different numbers of unit cells ($N$) of the SIP-CROW, evaluated exactly at the SIP frequency. The red solid curve is showing the fitted scaling of the group delay vs the length of the finite cavity as $aN^3 + b$, with $a = 32$ fs and $b = 5$ ps.

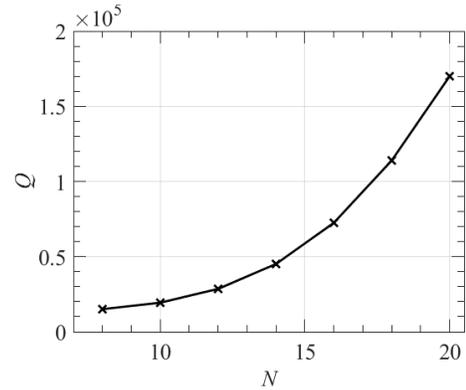

FIG.10. Quality factor of the CROW cavity shown in Fig. 1 calculated exactly at the SIP frequency for different number of unit cells ($N$) of the SIP-CROW.

As shown from Fig. 9, the group delay at the SIP frequency shows a scaling with the cavity length that is fitted by $aN^3 + b$, where the fitting parameters are $a = 32$ fs and $b = 5$ ps, and the asymptotic $N^3$ scaling was reported in [36]. Note that the group delay of a CROW cavity comprised of $N = 14$ unit cells is approximately equal to 74 ps, whereas the group delay of a single straight waveguide of the same length as the CROW is approximately equal to 2.7 ps. The value of the group delay of the single straight TL is calculated as $(L + 2L_{ex})/v_{ph}$, where the phase velocity $v_{ph} \simeq 1.27 \times 10^8$ m/s is calculated from the full-wave port analysis implemented in CST Studio Suite. This result shows that the group delay of the SIP-CROW made of $N = 14$ unit cells is 27 times larger than that of the single waveguide for this specific length, and this ratio gets higher for longer cavities. This clarifies the possible advantage of multimode waveguides operating near the SIP in delay line applications and for possible laser applications.

Finally, in Fig. 10, we show the quality factor calculated as $Q = \omega_e \tau_g / 2$ with $\omega_e$ being the SIP frequency [5], [24]. The quality factor is calculated for different numbers of unit cells, and it shows the $N^3$ scaling with cavity length, similar to the group delay. The proposed SIP-CROW could be used in applications that depend on compact optical delay lines.

## VII. Effect of Structural Perturbation on the SIP

We study the impact of perturbations in the modified CROW structure on the occurrence of the SIP. Indeed, during a microfabrication process, structural perturbations in the form of disorders and tolerances from the original design parameters occur. These disorders arise mainly from variations of the cross-sectional dimensions of the waveguides. These perturbations affect the effective modal refractive index as well as the coupling between the waveguides.

We mainly consider the perturbation that occurs for the waveguide width $w$. We assume that the fabrication imperfections in the waveguide width is due to etching and not due to mask imperfections. Thus, we assume that the waveguides widths are perturbed while preserving the alignment of the waveguides center axes. For the single waveguide portions, we assume that extensions $\Delta/2$ evenly occur on both sides of the waveguide as shown in Fig. 11(a). For the coupled waveguide portions, the perturbation results in width extensions on both sides of the waveguides and hence in narrowing the gap between the coupled waveguides.

We show in Fig. 11(c) the dispersion relation for the original design, with nominal waveguide width $w = w_0 = 450$ nm and gap $g = g_0 = 200$ nm, compared to the dispersion of the perturbed design with $\Delta = 10$ nm, which leads to $w = w_0 + \Delta = 460$ nm and $g = g_0 - \Delta = 190$ nm. Both curves show the SIP, with all the complex branches, i.e., the two evanescent branches and the almost purely real branch, coalescing at the SIP frequency. The dispersion shape with the occurrence of SIP is preserved in the perturbed case with extra width (red curves, right scale), however a frequency shift of about −1 THz occurs. Indeed, in Fig. 11(c) the dispersion diagram for the original design (blue curves, blue left scale) is plotted and compared with the dispersion diagram for the perturbed design (red curves, red right scale). Besides an overall frequency shift of about −1 THz occurs to the dispersion diagram relative to the $\Delta = 10$ nm width perturbation, the red and blue curves almost perfectly overlap. The dispersion diagrams shown in Fig. 11(c) are both based on CST full-wave simulations for one unit-cell, for each design.

We have also studied the effect of perturbations that may independently occur in reducing the gap (without changing the waveguide width), and in the silicon refractive index. We noticed negligible changes in the dispersion relation and to the occurrence of the SIP (without any significant frequency shift) when the gap is perturbed by ±20 nm or when the silicon refractive index is perturbed by ±0.03% (plots not shown for brevity).

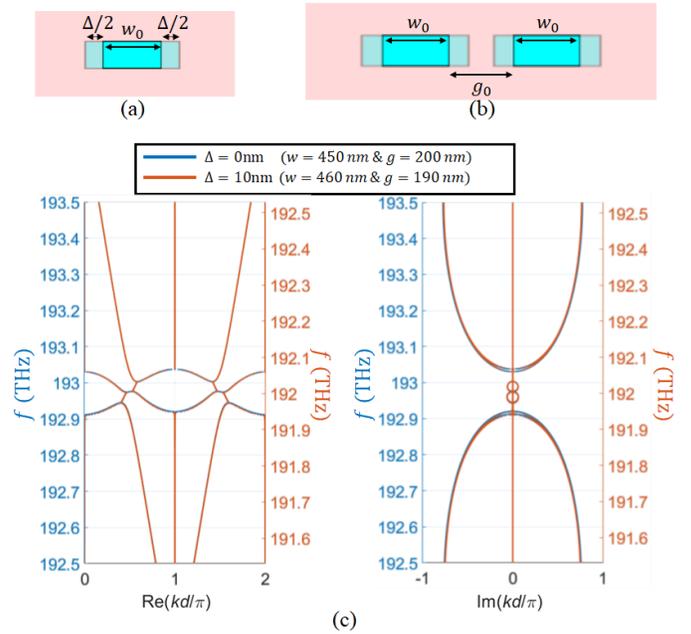

FIG. 11. Effect of width perturbation on the CROW waveguides. In (a) and (b) we show the fabrication perturbations affecting both the individual and coupled portions of the CROW. (c) Dispersion relation for the original design (blue, left scale) and perturbed design when $\mathbf{\Delta = 10}$ nm (red, right scale). The dispersion shape is preserved. By looking at the two scales, a frequency shift of about −1 THz occurs in the perturbed case; besides the shifted scales the red and blue curves overlap.

## VIII. Conclusion

We have provided a design of an SIP-CROW by utilizing distributed coupling through a directional coupler instead of the point coupling introduced in [5]. Also, we have introduced an extra degree of freedom by having two different radii in each ring resonator so that we can obtain the SIP with a compact unit cell consisting only of one ring (in [5] the SIP was obtained with a CROW with two rings in every unit cell). To facilitate the optimization of the unit cell dimensions and make it a realistic process, we used an integrated model based on full-wave simulations and an accurate transmission line model for the coupled directional coupler. We have illustrated the dispersion diagram showing an SIP at the optical wavelength 1550 nm. The dimensions of the unit cell are optimized using the concept of coalescence parameter. Finally, we have studied a finite length cavity operating in the vicinity of an SIP and we have shown that the transfer function is approximately unity at the SIP frequency without showing sharp resonances in the vicinity of the SIP. In addition, the finite length CROW operating at the SIP is showing very good matching that improves with using a greater number of unit cells. Moreover, we calculate the group delay and the quality factor of a finite length CROW operating at the SIP frequency and show that scale as $N^3$ with $N$ being the number of unit cells. The proposed CROW may find applications in sensors, lasers and optical delay lines.

## Acknowledgment

The authors are thankful to DS SIMULIA for providing CST Studio Suite that was instrumental in this study

## Appendix A: Simulation Setup for Mode Calculation

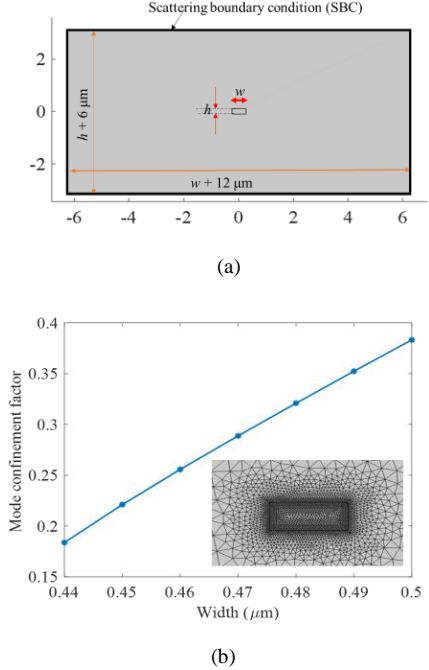

FIG.12. (a) Structure used for simulation. (b) Variation of mode confinement factor of the higher order TE mode with waveguide width, for fixed height of $h = 0.22$ μm. The mesh of the simulation is shown in inset.

The cross section of the simulation setup for mode calculations is shown in Fig.12(a). Here, and only for this task, we use COMSOL two-dimensional mode simulations using the Finite Element Method (FEM). The mesh is set with constraints that provides 10 nm mesh elements along the waveguide edges and a maximum mesh of 462 nm in the simulation region. Mode calculations are performed for a waveguide with height $h$ changing from 0.2 to 0.4μm, and width $w$ varied from 0.3 to 0.6μm. The height of the simulation domain is set to be 6 μm + $h$ (waveguide height) and the width of the simulation domain is set to be 12 μm + $w$ (waveguide width) as shown in Fig.12(a). The outer boundary of simulation domain is set as "scattering boundary condition". The mode confinement factor ($cf$), is calculated using the formula

$$cf = \frac{\int_{core}|\mathbf{H}|^2\,dx\,dy}{\int_{core\ and\ cladding}|\mathbf{H}|^2\,dx\,dy}. \quad (A1)$$

The confinement factor of the higher order TE mode is shown in Fig. 12(b) for a fixed waveguide height of $h = 220$ nm and variable waveguide width $w$. As the width is increased, the higher order TE mode gets more confined, and the calculated loss due to the scattering boundary condition is reduced. The higher TE mode in the waveguide with dimension of $450 \times 220$ nm has a confinement factor of 0.22.

## Appendix B: Definition of Equivalent Voltages and Currents Used in the Transfer Matrix Method

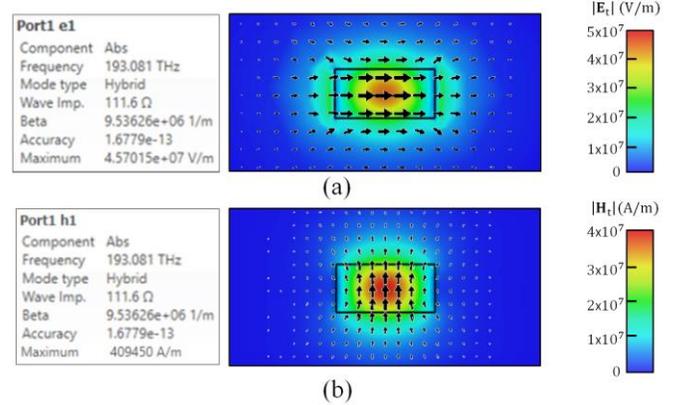

FIG.13. Mode profile in a single straight waveguide of cross-sectional dimensions 220 nm × 450 nm: (a) transverse electric field and (b) transverse magnetic field. The waveguide characteristic impedance $Z_{0w}$ and the propagation constant (called beta) are also shown at a frequency of 193.08 THz.

In this appendix, we show how the voltage and current are defined to model optical waveguides using TL. The used voltage and current are not physical ones, but they are equivalent ones that represent the electric and magnetic field in the structure. The way the voltage and current are defined in this paper is chosen to match the one used by the full-wave simulations implemented in CST Studio Suite.

The transverse electric and magnetic fields are represented in terms of equivalent voltage and current that describes the propagation along $z$ as

$$\begin{aligned}\mathbf{E}_t(x,y,z) &= V(z)\mathbf{e}(x,y),\\ \mathbf{H}_t(x,y,z) &= I(z)\mathbf{h}(x,y),\end{aligned} \quad (B1)$$

where $\mathbf{e}(x,y)$ and $\mathbf{h}(x,y)$ describe the transverse electric and magentic field mode profile in the transverse plane. The total power carried by the mode in the positive $z$-direction is [37], [38]

$$P = \frac{1}{2}\int_{-\infty}^{\infty}\int_{-\infty}^{\infty}\mathrm{Re}(\mathbf{E}_t \times \mathbf{H}_t^* \cdot \hat{\mathbf{z}})dxdy = \frac{1}{2}\mathrm{Re}(VI^*). \quad (B2)$$

We show in Fig. 13 the transverse electric and magentic field profile, $\mathbf{E}_t$ and $\mathbf{H}_t$, for a wave that carries power of 1/2 Watt in $z$-direction.

The amplitudes of the incident and reflected power waves [$\sqrt{W}$] are defined in terms of the voltage and current as [37], [38]

$$\begin{aligned}a &= \frac{1}{2\sqrt{Z_{0w}}}(V + Z_{0w}I),\\ b &= \frac{1}{2\sqrt{Z_{0w}}}(V - Z_{0w}I).\end{aligned} \quad (B2)$$

The port analysis implemented in CST Studio Suite characterizes the mode using two parameters: the wavenumber $k_w$ that describes the phase propagation of the mode, and the wave impedance $Z_{0w}$ that is calculated by CST as the average of the ratios between the transverse electric field $\mathbf{E}_t$ and magnetic field $\mathbf{H}_t$ across the transverse plane. An example of the value of the wavenumber $k_w$ (Beta in the legend) and the wave impedance calculated around 193 THz are shown in the legend of Fig. 13.

Similarly, for two uniform *coupled* optical waveguides, the voltages and currents that represent the even and odd modes are defined as

$$\mathbf{E}_{t,e}(x, y, z) = V_e(z)\mathbf{e}_e(x, y),$$
$$\mathbf{H}_{t,e}(x, y, z) = I_e(z)\mathbf{h}_e(x, y),$$
$$\mathbf{E}_{t,o}(x, y, z) = V_o(z)\mathbf{e}_o(x, y),$$
$$\mathbf{H}_{t,o}(x, y, z) = I_o(z)\mathbf{h}_o(x, y).$$
(B3)

where $Z_{0e}$ and $Z_{0o}$ are the wave impedances of the even and odd modes, respectively, and $\mathbf{e}_e$, $\mathbf{e}_o$, $\mathbf{h}_e$ and $\mathbf{h}_o$ describe the transverse electric and magnetic fields for even and odd modes. We show in Fig. 14 the transverse electric field for even and odd modes.

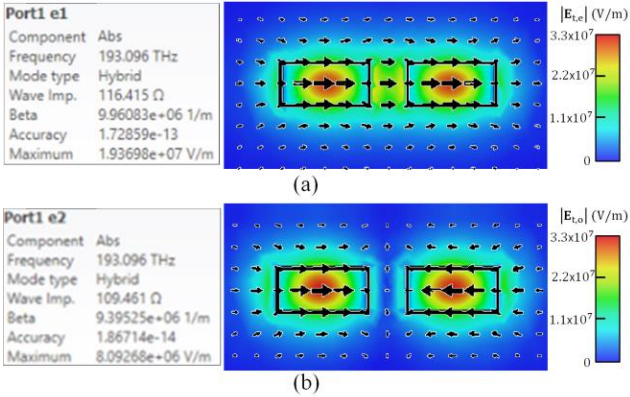

FIG.14. (a) and (b) are the mode profiles of the even and odd mode, respectively, in two coupled straight waveguides cross sectional dimensions 220 nm × 450 nm, where the gap size between the coupled waveguide $g$ = 200 nm. The wave impedances $Z_{0e}$ and $Z_{0o}$ together with the even and odd propagation constants $k_e$ and $k_o$ are shown. These results are obtained using full-wave simulations based on the finite element method implemented in CST Studio Suite.

The power wave amplitudes are expressed in term of the voltage and the current for even and odd modes as

$$a_e = \frac{1}{2\sqrt{Z_{0e}}}(V_e + Z_{0e}I_e), \quad b_e = \frac{1}{2\sqrt{Z_{0e}}}(V_e - Z_{0e}I_e),$$
$$a_o = \frac{1}{2\sqrt{Z_{0o}}}(V_o + Z_{0o}I_o), \quad b_o = \frac{1}{2\sqrt{Z_{0o}}}(V_o - Z_{0o}I_o).$$
(B4)

In the following appendix we show how to build the transfer matrices for uniform waveguides section based on a state vector definition that involves voltages and currents defined in this appendix. In Appendix D, we show how that S-parameters (that are obtained based on power wave amplitudes defined in this appendix) are used to obtain the transfer matrices for blocks with curved waveguides.

## Appendix C: Transmission Line Model of Coupled and Uncoupled Waveguide

In this appendix, we present the transfer matrices of the transmission line model used for coupled and uncoupled waveguide sections of the presented unit cell in Fig. 2(a). The whole transfer matrix that describes the relation between the fields at the begin and end of the unit cell is calculated by cascading different transfer matrices. Care should be taken when we merge matrices with different dimensions, being also sure that we apply consistent definitions for the state vectors at different cross-sections (it is easy to make a sign error when connecting two different blocks).

The transfer matrices of the uniform blocks in the unit cell in Fig. 2(a) are used for either a single waveguide described by 2×2 matrix, or two coupled waveguides described by 4×4 matrix. For the sake of brevity, we do not present the details of cascading the T-matrices of the different subblocks. The presented transfer matrices in this appendix describe the evaluation of the state vector that involve voltages and current that are defined in Appendix B.

*Single uncoupled transmission line* (TL): the state vector that represents single mode propagation in the waveguide is defined as

$$\mathbf{\Psi}(z) = [V(z), \quad I(z)]^T. \qquad (C1)$$

The T-matrix describes the propagation of the modes in a waveguide with length $L_w$ as $\mathbf{\Psi}(z + L_w) = \underline{\mathbf{T}}\mathbf{\Psi}(z)$, and it is given by

$$\underline{\mathbf{T}} = \begin{bmatrix} \cos(k_w L_w) & -jZ_{0w}\sin(k_w L_w) \\ -j\sin(k_w L_w)/Z_{0w} & \cos(k_w L_w) \end{bmatrix}, (C2)$$

where $k_w = \omega/v_{ph}$ is the propagation constant with $v_{ph}$ being the guided phase velocity of the propagating wave, and $Z_{0w}$ is the characteristic impedance of the TL. The value of the phase velocity and characteristic impedance are calculated from the full-wave port analysis implemented in CST Studio Suite.

The characteristic impedance used in the in the T matrix in Eq. (C2) is assumed to have the same value of the wave impedance found by the 2D full-wave port analysis implemented in CST Studio Suite, defined as discussed in Appendix B. The modal wavenumber used in Eq. (C2) is the same as the one estimated by the port analysis.

*Coupled transmission lines (CTL)*: The T-matrix that describes the propagation of the modes in CTLs of length $L_C$ is better defined in terms of the independent even and odd modes. The state vector representing the even and odd voltages and currents in that case is written as

$$\mathbf{\Psi}_{eo}(z) = [V_e(z), \quad I_e(z), \quad V_o(z), \quad I_o(z)]^T \quad (C3)$$

and in that case the T-matrix describing the evolution, $\mathbf{\Psi}_{eo}(z + L_C) = \underline{\mathbf{T}}_{eo}\mathbf{\Psi}_{eo}(z)$, of the odd modes is given by

$$\mathbf{T}_{eo} = \begin{bmatrix} \cos(k_e L_C) & -jZ_{0e}\sin(k_e L_C) & 0 & 0 \\ \dfrac{-j\sin(k_e L_C)}{Z_{0e}} & \cos(k_e L_C) & 0 & 0 \\ 0 & 0 & \cos(k_o L_C) & -jZ_{0o}\sin(k_o L_C) \\ 0 & 0 & \dfrac{-j\sin(k_o L_C)}{Z_{0o}} & \cos(k_o L_C) \end{bmatrix} \quad (C4)$$

where $k_e$ and $k_o$ are the propagation constants of the even and odd modes given by $k_e = \omega/v_{ph,e}$ and $k_o = \omega/v_{ph,o}$, respectively, with $v_{ph,e}$ and $v_{ph,o}$ being the guided phase velocities of the even and odd modes, respectively. $Z_{0e}$ and $Z_{0o}$ are the wave impedances of the even and odd modes. In this paper, we find the fundamental CTL parameters $v_{ph,e}$, $v_{ph,o}$, $Z_{0e}$, and $Z_{0o}$ directly from full-wave port-analyses simulations. The T matrix in (C4) is evaluated using these parameters. The length of the CTL is determined based on an optimizer for obtaining the SIP as explained in the body of the paper.

For the *single waveguide* made from silicon ($n_{si}$=3.48), with dimensions of $450 \times 220$ nm and with a cladding made of silicon dioxide ($n_c$=1.45), we found from the port analysis finite element method implemented in CST Studio Suite simulations that the impedance and phase velocity calculated at $f = 193.08$ THz are $v_{ph} = 0.424c$ and $Z_{0w} = 111.6$ Ohm, respectively, where $c$ is free space speed of light.

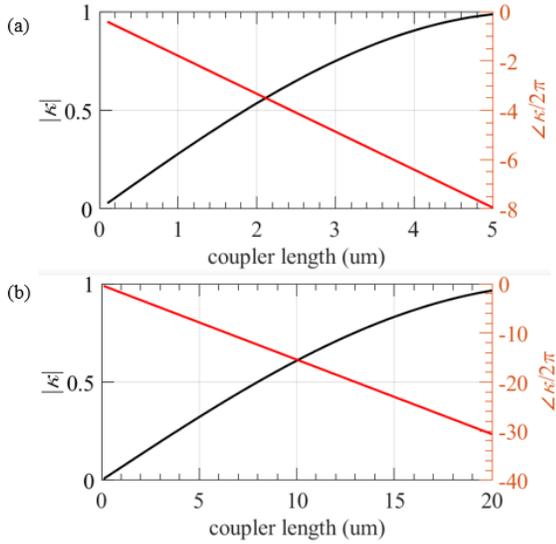

FIG.15. (a) Magnitude and phase of the field coupling coefficient $\kappa$ versus length of a directional coupler comprised of two coupled waveguides with a gap size 80 nm. Unitary coupling is achieved in 5 μm. (b) Magnitude and phase of the field coupling coefficient versus the length of the directional coupler in the case of 200 nm gap size. Unitary coupling is achieved in 20 μm. Note that for a directional coupler of length $L_{rr} = 0.365$ μm, the coupling coefficient between the rings is approximately 2.4%. It is 1.8% between the rings and the straight waveguide for a coupler length of $L_{wr} = 0.273$ μm.

For the *coupled waveguides* with gap of 200 nm, the port analysis carried out at $f = 193.09$ THz by using CST Studio Suite leads to the even and odd modes' phase velocities and impedances $v_{ph,e} = 0.412c$, $v_{ph,o} = 0.427c$, $Z_{0e} = 112.43$ Ohm and $Z_{0o} = $ 110.87 Ohm. We used an adaptive mesh with stopping criterion of threshold $5 \times 10^{-5}$ for the error of port mode normalized wavenumber $|k_z/k_0|$.

The field coupling coefficient is dictated by the length of the directional coupler and the gap between the two coupled waveguide. In Fig. 15, we show the magnitude and phase of the field coupling coefficient $\kappa$ and the field transmission coefficient for two gap sizes, 80 nm (shown in Fig. 15(a)) and 200 nm (shown in Fig. 15(b)). From these two figures we see that an almost 100% coupling can be achieved using only 5 μm couplers with gap 80 nm, while we need 20 μm in the case of 200 nm. The magnitude of the field transmission $\tau$ is related to $\kappa$ as

$$|\tau|^2 + |\kappa|^2 \simeq 1, \quad (C5)$$

In the design proposed in this paper, the magnitude of the field coupling coefficient between the rings is approximately 2.4% which corresponds to a directional coupler of length $L_{rr}$=0.365 μm. While it is 1.8% between each ring and the straight waveguide corresponding to a directional coupler of length $L_{wr} = 0.273$ μm.

**Appendix D: Curved Waveguides S-parameters-Based Model**

In this appendix, we present the method we used to find the transfer matrices of the curved waveguides blocks highlighted by blue color in Fig. 2(a). For sake of brevity, we show the procedure for one of the subblocks since it is the same for other blocks.

We obtain the scattering matrix for the sub-block shown in Fig. 16(a) using full-wave simulations implemented in CST Studio Suite. We connected the two coupled waveguides at the beginning with one port, Port1, that excites 2 modes, even and odd modes. We connected the end of the structure with two ports, Port2 and Port3, as illustrated in Fig. 16(a), where a single mode is defined for each of Port2 and Port3. For numerical simplicity, when using commercial simulators like CST Studio Suite, we obtain the scattering parameters referred to impedance of $Z_0 = 50$ Ohm. This used reference impedance $Z_0$ should be cancelled out when we calculate the T matrix.

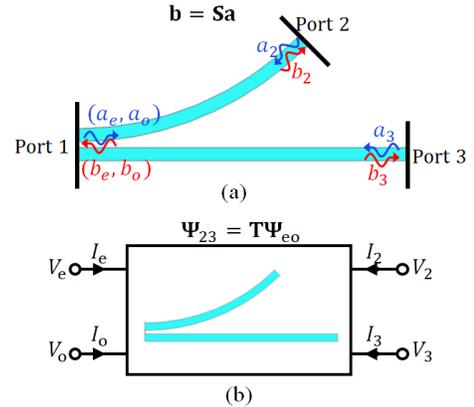

FIG.16. (a) One of the sub-blocks with curved waveguide showing incident and scattered waves upon which the S-matrix is calculated. (b) Equivalent block model in term of voltage and current found by transforming the scattering matrix to transfer matrix.

The S-matrix for the sub-block shown in Fig. 16(a) give the relation between the incident and reflected power waves as $\mathbf{b} = \underline{\mathbf{S}}\,\mathbf{a}$, where

$$\mathbf{a} = [a'_e, \ a'_o, \ a'_2, \ a'_3]^T,$$
$$\mathbf{b} = [b'_e, \ b'_o, \ b'_2, \ b'_3]^T. \quad (D1)$$

The power wave amplitudes in (D1) are defined as

$$a'_e = \frac{1}{2\sqrt{Z_0}}(V_e + Z_0 I_e), \quad b'_e = \frac{1}{2\sqrt{Z_0}}(V_e - Z_0 I_e),$$
$$a'_o = \frac{1}{2\sqrt{Z_0}}(V_o + Z_0 I_o), \quad b'_o = \frac{1}{2\sqrt{Z_0}}(V_o - Z_0 I_o),$$
$$a'_2 = \frac{1}{2\sqrt{Z_0}}(V_2 + Z_0 I_2), \quad b'_2 = \frac{1}{2\sqrt{Z_0}}(V_2 - Z_0 I_2),$$
$$a'_3 = \frac{1}{2\sqrt{Z_0}}(V_3 + Z_0 I_3), \quad b_3 = \frac{1}{2\sqrt{Z_0}}(V_3(z) - Z_0 I_3), \quad (D2)$$

and the equivalent voltages and currents are similarly defined as in Appendix B.

Note that the power wave amplitudes in (D2) are different from the ones defined in Appendix B because the ones in (D2) are defined with respect to a reference impedance $Z_0 = 50$ Ohm, whereas the ones in Appendix B are defined with respect to the wave impedance that describes each mode.

To integrate the subblocks model with the model for uniform sections presented in Appendix B, the state vector should be consistently defined at the interfaces between subblocks, i.e., should be defined based on voltages and currents.

We convert the scattering matrix calculated using the full-wave solver to the transfer matrix that describes the voltages and currents at the beginning and end of the subblock, as schematically shown in Fig. 16(b). The T-matrix for this case describes the state vectors as $\mathbf{\Psi}_{23} = \underline{\mathbf{T}}\,\mathbf{\Psi}_{eo}$, where

$$\mathbf{\Psi}_{23} = [V_2, \ -I_2, \ V_3, \ -I_3]^T,$$
$$\mathbf{\Psi}_{eo} = [V_e, \ I_e, \ V_o, \ I_o]^T. \quad (D3)$$

The transformation from $\underline{\mathbf{S}}$ matrix to $\underline{\mathbf{T}}$ matrix is performed based on the relations given in Eq. (D2).

We rewrite Eq (D2) in matrix form as

$$\mathbf{a} = \underline{\mathbf{t}}_1 \mathbf{\Psi}_{eo} + \underline{\mathbf{t}}_2 \mathbf{\Psi}_{23},$$
$$\mathbf{b} = \underline{\mathbf{t}}_3 \mathbf{\Psi}_{eo} + \underline{\mathbf{t}}_4 \mathbf{\Psi}_{23}, \quad (D4)$$

where

$$\underline{\mathbf{t}}_1 = \frac{1}{2\sqrt{Z_0}}\begin{bmatrix} 1 & Z_0 & 0 & 0 \\ 0 & 0 & 1 & Z_0 \\ 0 & 0 & 0 & 0 \\ 0 & 0 & 0 & 0 \end{bmatrix}, \quad \underline{\mathbf{t}}_2 = \begin{bmatrix} 0 & 0 & 0 & 0 \\ 0 & 0 & 0 & 0 \\ 1 & Z_0 & 0 & 0 \\ 0 & 0 & 1 & Z_0 \end{bmatrix},$$
$$\underline{\mathbf{t}}_3 = \frac{1}{2\sqrt{Z_0}}\begin{bmatrix} 1 & -Z_0 & 0 & 0 \\ 0 & 0 & 1 & -Z_0 \\ 0 & 0 & 0 & 0 \\ 0 & 0 & 0 & 0 \end{bmatrix}, \quad \underline{\mathbf{t}}_4 = \begin{bmatrix} 0 & 0 & 0 & 0 \\ 0 & 0 & 0 & 0 \\ 1 & -Z_0 & 0 & 0 \\ 0 & 0 & 1 & -Z_0 \end{bmatrix}. \quad (D5)$$

Substituting (D5) in $\mathbf{b} = \underline{\mathbf{S}}\,\mathbf{a}$, the transfer matrix is found in terms of the scattering parameters matrix as

$$\underline{\mathbf{T}} = (\underline{\mathbf{t}}_4 - \underline{\mathbf{S}}\,\underline{\mathbf{t}}_2)^{-1}(\underline{\mathbf{S}}\,\underline{\mathbf{t}}_1 - \underline{\mathbf{t}}_3). \quad (D6)$$

The transformation matrices in (D5) and the scattering parameters matrix $\underline{\mathbf{S}}$ depend on the reference impedance $Z_0$, but the mathematical simplification of the transfer matrix in (D6) yields a matrix that is independent of the reference impedance $Z_0$. Therefore, the choice of the reference impedance $Z_0$ does not affect the obtained transfer matrix $\underline{\mathbf{T}}$. This makes sense because the transfer matrix $\underline{\mathbf{T}}$ characterizes the subblock without including the reference impedance at the ports, i.e., the relation between voltages and currents at beginning and the end of subblock is not governed by the reference impedance at the ports.

**Appendix E: Validation of the Hybrid Model Based on Full-wave Simulations and TL-Coupler Model**

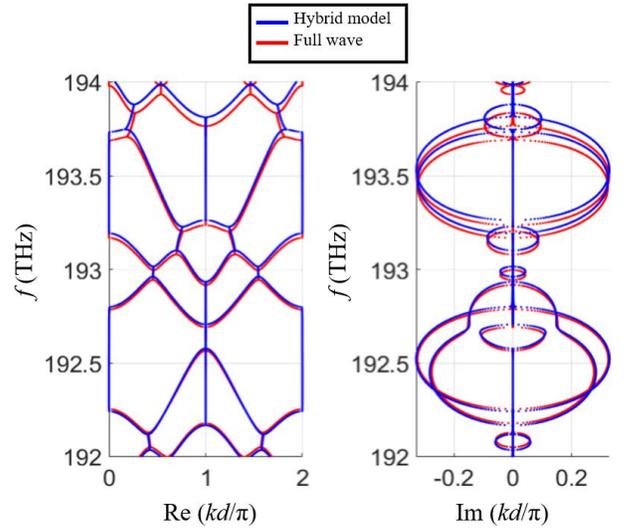

FIG.17. Comparison of the complex dispersion diagram obtained using the hybrid model vs that obtained from full-wave simulations based on the finite element method implemented in CST studio suite. The one-ring unit cell has the dimensions $g = 80$ nm, $R_t = 10$ μm, $R_b = 11.95$ μm, $L_{rr} = 1.12$ μm, and $L_{wr} = 0.35$ μm. The two models have an excellent agreement.

In Fig. 17, we compare the dispersion diagram obtained from the hybrid model to the dispersion diagram based completely on the full-wave simulations implemented in CST Studio Suite. The hybrid model uses T-matrices based on TL equations to model the coupled sections of the unit cell, whereas the T-matrices of the other segments in the unit cell are obtained from full-wave simulations (see Fig. 2). The two dispersion diagrams are in a very good agreement, which validates the proposed hybrid model. The proposed structure with the dimensions used in this section does not exhibit an SIP. However, the structure is very close to exhibit an SIP because the dispersion diagram shown in Fig. 17 shows three modes with wavenumbers close to each other at 193 THz, the structure exhibits so-called titled SIP. The comparison between the two models when the system exhibits an SIP is shown in Fig. 5.